\documentclass[12pt]{article}
\usepackage{graphicx}
\usepackage{subfigure}

\newcommand\pubnumber{}
\newcommand\pubdate{\today}

\def\gucasadd{Graduate University of Chinese Academy of Sciences\\
19A Yuquan Road, Beijing, 100049, China \\
zhengyh$@$gucas$.$ac$.$cn}
\def\support{\footnote{Work supported by the Ministry of Science
and Technology of China, National Natural Science Foundation of
China and Chinese Academy of Sciences}}

\def\Title#1{\begin{center} {\Large #1 } \end{center}}
\def\Author#1{\begin{center}{ \sc #1} \end{center}}
\def\Address#1{\begin{center}{ \it #1} \end{center}}

\newcommand\pubblock{\rightline{\begin{tabular}{l} \pubnumber\\
         \pubdate  \end{tabular}}}
\newenvironment{Abstract}{\begin{quotation}  }{\end{quotation}}
\newenvironment{Presented}{\begin{quotation} \begin{center}
             PRESENTED AT\end{center}\bigskip
      \begin{center}\begin{large}}{\end{large}\end{center} \end{quotation}}
\def\Acknowledgements{\bigskip  \bigskip \begin{center} \begin{large}
             \bf ACKNOWLEDGEMENTS \end{large}\end{center}}

\def\beq{\begin{equation}}
\def\eeq#1{\label{#1}\end{equation}}
\def\eeqn{\end{equation}}

\def\beqa{\begin{eqnarray}}
\def\eeqa#1{\label{#1}\end{eqnarray}}
\def\eeqan{\end{eqnarray}}

\let\bar=\overbar

\def\Dslash{\not{\hbox{\kern-4pt $D$}}}
\def\dslash{\not{\hbox{\kern-2pt $\del$}}}

\def\msb{{\bar{\ssstyle M \kern -1pt S}}}

\begin{document}
\begin{titlepage}
\pubblock

\vfill \Title{Current and Future Charm Experiments} \vfill \Author{
Yangheng Zheng\support} \Address{\gucasadd} \vfill
\begin{Abstract}
Charm physics has been studied with many dedicated accelerators and
experiments. In these proceedings, we review the status and the
selected results of the current operating BEPCII/BESIII experiment.
We also discuss the BESIII data taking plan for the future.
\end{Abstract}
\vfill
\begin{Presented}
The 6th International Workshop on the CKM Unitarity Triangle\\
University of Warwick, Coventry, United Kingdom,  September 6--10,
2010
\end{Presented}
\vfill
\end{titlepage}
\def\thefootnote{\fnsymbol{footnote}}
\setcounter{footnote}{0}

\section{Introduction}
\begin{figure}[htb]
\centering
\includegraphics[height=1.8in]{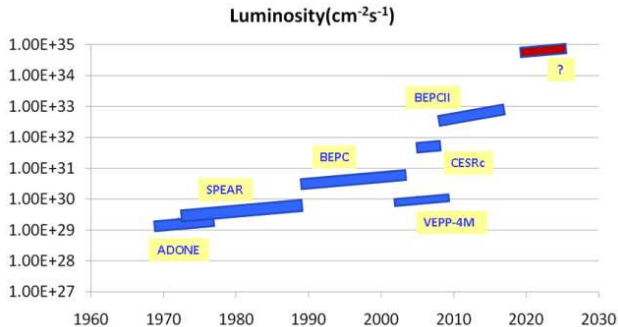}
\caption{Dedicated $e^+e^-$ colliders for charm physics}
\label{fig:eecolliders}
\end{figure}

Since the 1960's, many electron-positron colliders have been built
around the tau-charm energy regions, with a clear tendency of
luminosity increasing in time (see Figure~\ref{fig:eecolliders}).
The most renowned of these colliders was SPEAR, which discovered
both the $J/\psi$ and $\tau$, and began the era of $\tau$-charm
physics.

In the early 1990's, a precise measurement facility, named BEPC/BES,
was built at the $\tau$-charm energy region in IHEP, China. It is
the first accelerator built for high energy physics research in
China. The luminosity was an order of magnitude higher than SPEAR.
In the year 2004, CESR/CLEO decreased its collision energy to the
charm energy region to start the  CESRc/CLEOc programme, and became
the highest luminosity charm facility then existing.

BEPCII/BESIII is a mainstream high energy physics project in
China~\cite{Yifang, BESIIINIM}. It is an upgrade from its ancestor
BEPC/BES. The designed peak luminosity is $1 \times 10^{33} \
\mathrm{cm^{-2} s^{-1}}$, which is an order of magnitude higher than
CESRc/CLEOc. BEPCII/BESIII started its test run in 2008 and took
physics data in 2009. By the end of 2010, many physics results had
already been published in journals.

Therefore, there have existed dedicated charm experiments for over
40 years. There is no doubt that this is a very interesting energy
region with rich physics in it. The charm quark can be seen as a
bridge which links perturbative QCD and non-perturbative QCD, and
studies in charm physics can provide calibrations and tests of
Lattice QCD. With high statistics production of charm mesons at
threshold, we can precisely measure the CKM matrix elements $V_{cd}$
and $V_{cs}$, the absolute branching fractions of charm meson
decays, the decay constants $f_{D^+}$ and $f_{D_s}$, light meson
spectroscopy in Dalitz plot analyses etc. We can also probe neutral
charm meson mixing, search for $CP$ violation, and measure strong
phase differences with the quantum correlation of the charm meson
pairs.

Charm physics is a topic of interest at hadron colliders, i.e. the
Tevatron and LHC. They take advantage of the huge charm production
cross-section and the high energy boost of the charm meson decay
vertices. However, the $e^+e^-$ experiments have much cleaner
collision environment with almost 100\% trigger efficiency. The
quantum correlations and meson tagging techniques can be applied to
the threshold production experiments, i.e. CLEOc and BESIII. There,
many systematic uncertainties can be cancelled out while applying
the "double tag" method which also leads to very pure samples.

\section{BEPCII and BESIII performance}
BEPCII is a double-ring $e^+e^-$ collider with a designed peak
luminosity of $10^{33} \ \mathrm{cm^{-2} s^{-1}}$ at a beam current
of $0.93$ A. It covers the cms energy region from 2.0 GeV to 4.6
GeV. During the 2009 to 2011 operating period, both the accelerator
and the detector performed remarkably well and the world largest
data sample of $J/\psi$, $\psi$' and $\psi(3770)$ have been
collected at threshold energy. This impressive early performance
indicates that the experiment has great potential for improving our
understanding of the physics that can be accessed in the
$\tau$-charm regime. A short summary of the BEPCII/BESIII status and
performance will be given below.

The BEPCII/BESIII upgrade project started in 2003 and successfully
completed in 2008. BEPCII managed to accumulate a beam current of
$500$ mA in the storage ring, and obtained a collision luminosity
close to $10^{32} \ \mathrm{cm^{-2} s^{-1}}$ in March 2008.
Installation of the BESIII detector was completed at the end of 2007
and the first full cosmic-ray event was recorded in March 2008. The
detector was successfully moved to the interaction point on April
30, 2008. With a careful tuning of the machine, the first $e^+e^-$
collision event was recorded by the BESIII detector on July 19,
2008. In the test run period, a total of 14 million $\psi$' events
was collected until Nov. 2008. Over this period, the BEPCII
performance continued to improve through lattice optimization,
system debugging, and vacuum improvements. After a $1.5$-month
synchrotron radiation run and a winter maintenance, the machine
resumed collision and its luminosity gradually improved to $3 \times
10^{32} \ \mathrm{cm^{-2} s^{-1}}$.

The official data taking for physics run started in March 2009. In
this first run period, BESIII successfully collected 100 million
$\psi$' events and $200$ million $J/\psi$ events, about a factor of
4 larger than the previous data samples from CLEOc and BESII,
respectively. The peak luminosity was stable, typically at the level
of $2 \times 10^{32} \ \mathrm{cm^{-2} s^{-1}}$ during the data
taking at $\psi$', and $0.6 \times 10^{32} \ \mathrm{cm^{-2}
s^{-1}}$ at $J/\psi$. An energy scan of the $\psi$' line-shape shows
that the beam energy spread is about $1.4$ MeV, and the effective
peak cross section of $\psi$' production is about $700$ nb.

Starting from December 2009, BESIII collected data around the
$\psi(3770)$ resonance. Up to the time of this write-up, more than
$1 ~ \mathrm{fb^{-1}}$ data was recorded, which included
$75~\mathrm{pb^{-1}}$ data for an energy scan of the $\psi(3770)$
line-shape. During the 2010 to 2011 run period, a peak luminosity of
$5.6 \times 10^{32} \ \mathrm{cm^{-2} s^{-1}}$ was achieved. The
data taking efficiency of the detector is more than $85\%$.

In common with many other general purpose detector for high energy
physics, the cylindrical core of the BESIII detector consists of a
helium-gas-based drift chamber (MDC), a plastic scintillator
Time-of-Flight system (TOF), and a CsI(Tl) Electromagnetic
Calorimeter (EMC), all enclosed in a superconducting solenoidal
magnet providing a $1.0$ T magnetic field. The solenoid is supported
by an octagonal flux-return yoke with resistive plate counter muon
identifier modules (MU) interleaved with steel. The charged particle
and photon acceptance is $93\%$ of $4\pi$, and the charged particle
momentum and photon energy resolutions at $1$ GeV are $0.5\%$ and
$2.5\%$, respectively.

A comprehensive Monte Carlo simulation software package, largely
based on the first principle of particles interacting with detector
materials, was developed to model the performance of the BESIII
detector. Comparing with the data, good agreement was observed, not
only on average numbers, but also in the shape of many
distributions. Data analysis shows that the detector is in a very
good condition and all the design specifications have been
satisfied.

\section{Selected result}
The physics programme as BESIII is very rich~\cite{BESIIIYellow}.
Here, we just give some examples from the first published
results~\cite{BESIIIHc,BESIIIChicJ2PP,BESIIIGammaPP} to show the
potential for future precise and interesting measurements. These
results ranged from the confirmation of BESII and CLEOc results, to
completely new observations.

\begin{figure}[htb]
\centering

{\includegraphics[width=2in]{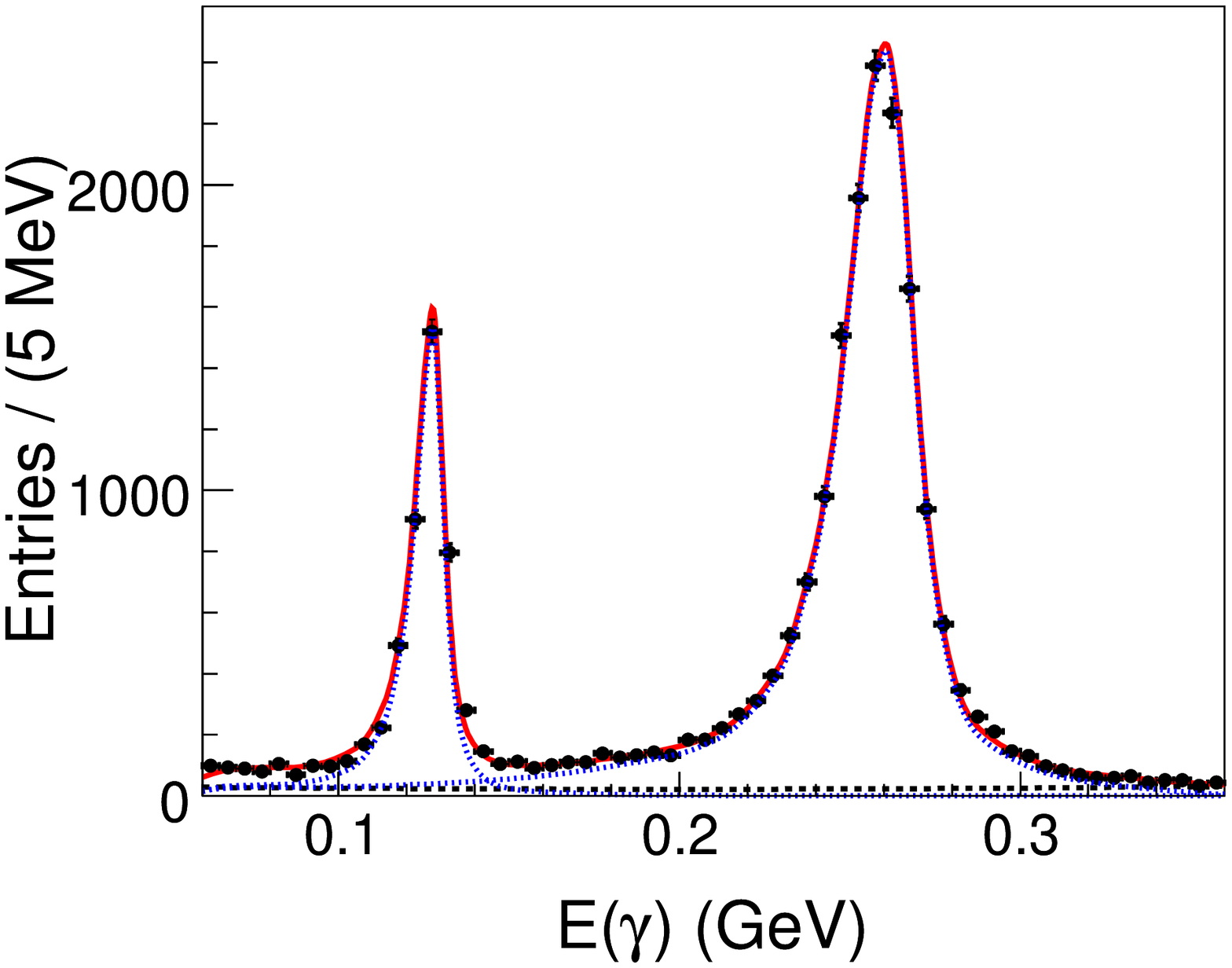}}
{\includegraphics[width=2in]{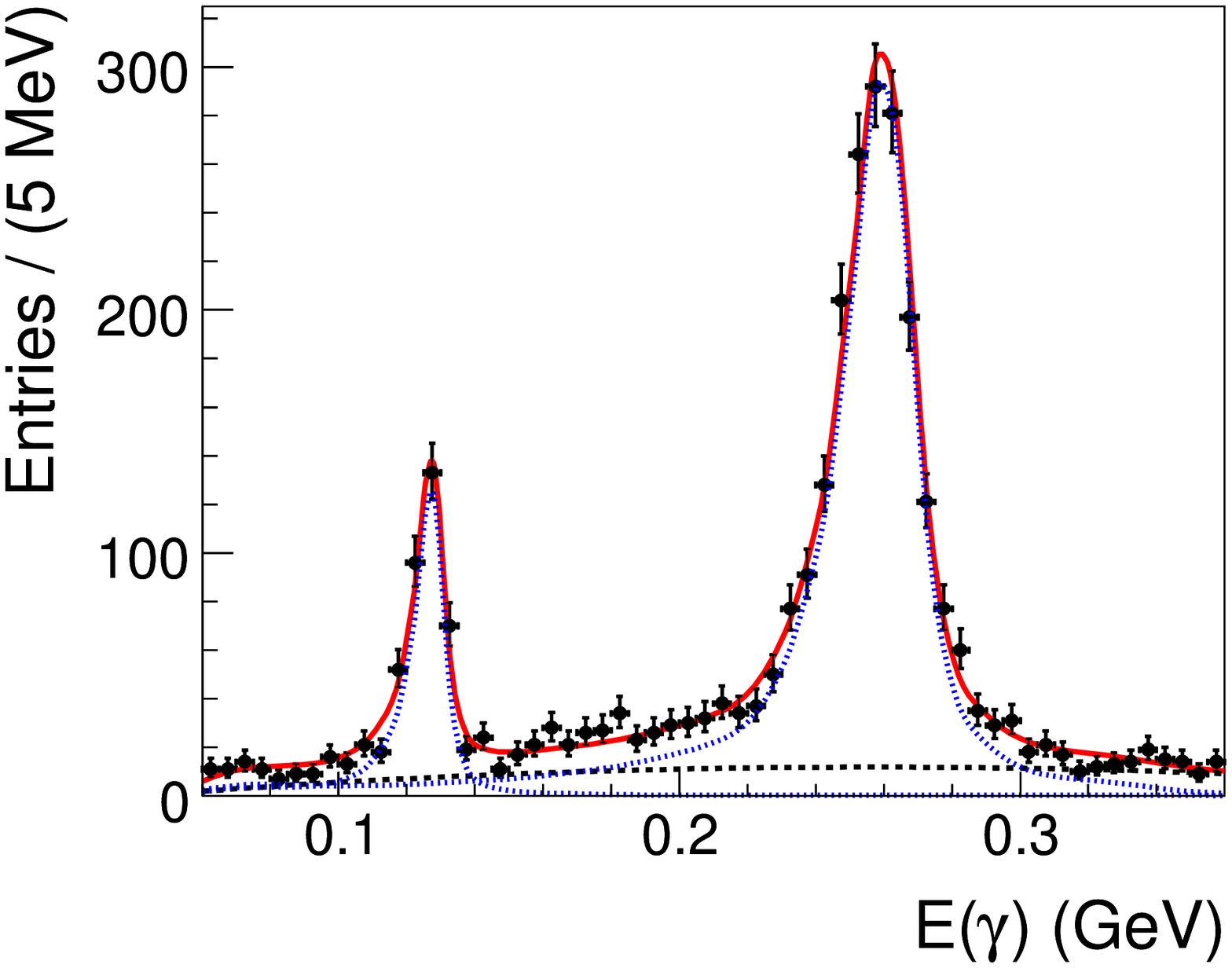}}

\caption{The radiative photon energy spectrum of $\chi_{c0}$ and
$\chi_{c2}$ signals. Left plot is from selected $\psi\prime
\rightarrow \gamma \pi^0 \pi^0$ events, and right plot is from
selected $\psi\prime \rightarrow \gamma \eta \eta$ events.}
\label{fig:chic2PP}
\end{figure}

Figure~\ref{fig:chic2PP} shows the prompt photon spectrum from
$\psi\prime \rightarrow \gamma \pi^0 \pi^0$ and $\psi\prime
\rightarrow \gamma \eta \eta$ channels. Signals of $\chi_{c0}$ and
$\chi_{c2}$ are observed and their branching ratios are measured
~\cite{BESIIIChicJ2PP}. The results are consistent with the CLEOc
measurements\cite{CLEOc3}.

\begin{figure}[htb]
\centering
\includegraphics[height=2in]{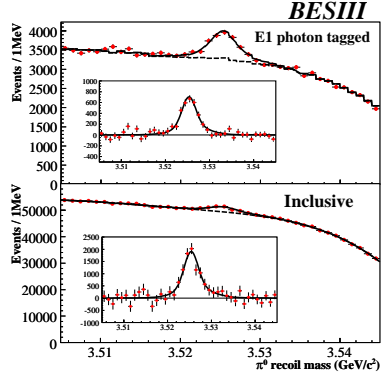}
\caption{The $\pi^0$ recoil mass spectrum and fit. Upper: tagged
with the prompt photon in the $h_c \rightarrow \gamma \eta_c$ decays
($E1$-tagged analysis).  Lower: inclusive analysis of $\psi\prime
\rightarrow \pi^0 h_c$. } \label{fig:hsubc}
\end{figure}

The last member of the charmoniuum family below the open charm
threshold, called the $h_c$, was observed by CLEOc in 2005 from
$\psi$' decays to $\pi^0 h_c$, $h_c \rightarrow \gamma
\eta_c$~\cite{CLEOc0} and an improved measurement was performed
recently~\cite{CLEOc1}. BESIII performed a similar analysis with a
larger data sample, and a clear signal can be seen by tagging the
prompt photon in the $h_c$ decays~\cite{BESIIIHc}, as shown in
Figure~\ref{fig:hsubc}. In addition, BESIII looked for inclusive
$\pi^0$ production from $\psi$' decays and clear signals can be also
seen. Thus, the branching fractions of $\psi\prime \rightarrow \pi^0
h_c$, $h_c \rightarrow \gamma \eta_c$ can be individually measured
for the first time, together with the width of $h_c$. The results
are listed in Table~\ref{tab:results} in comparison with recent CLEO
results~\cite{CLEOc1}. Good agreement can be seen.
\begin{table}[t]
\begin{center}
\begin{tabular}{|l|c|c|}
\hline Measured parameters &  BESIII results &  CLEOc results
\\ \hline
 Mass: $M_{h_c}$ & $3525.40 \pm 0.13 \pm 0.18$ MeV &
 $3525.28 \pm 0.19 \pm 0.12$ MeV \\ \hline
 Width: $\Gamma_{h_c}$ & $0.73 \pm 0.45 \pm 0.28$ MeV & - \\ \hline
 $B(\psi\prime \rightarrow \pi^0 h_c)$ & $(8.4 \pm 1.3 \pm 1.0) \times 10^{-4}$  &
  - \\ \hline
 $B(h_c \rightarrow \gamma \eta_c)$ & $(54.3 \pm 6.7 \pm 5.2)\%$ & - \\ \hline
 $B(\psi\prime \rightarrow \pi^0 h_c) \times B(h_c \rightarrow \gamma \eta_c)$ &
 $(4.58 \pm 0.40 \pm 0.50) \times 10^{-4}$ &
 $(4.19 \pm 0.32 \pm 0.45) \times 10^{-4}$ \\
 \hline
\end{tabular}
\caption{Measured results in comparison with CLEOc. (In the fit,
$\Gamma_{h_c}$ is floated in the BESIII analysis while CLEOc fixes
$\Gamma_{h_c} = \Gamma_{\chi_{c1}} = 0.9$ MeV)} \label{tab:results}
\end{center}
\end{table}

\begin{figure}[htb]
\centering
\includegraphics[width=2in]{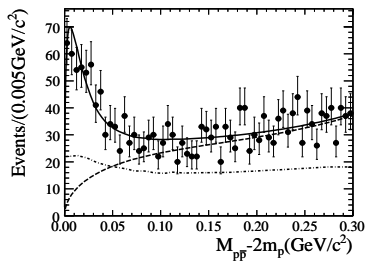}
\caption{The $p\bar{p}$ invariant mass spectrum for the $\psi\prime
\rightarrow \pi^+ \pi^- J/\psi \ (J/\psi \rightarrow \gamma p
\bar{p})$ after final event selection. The solid curve is the fit
result; the dashed curve shows the fitted background function, and
the dash-dotted curve indicates how the acceptance varies with
$p\bar{p}$ invariant mass} \label{fig:BESIIIgammapp}
\end{figure}

The $p\bar{p}$ threshold enhancement in $J/\psi \rightarrow \gamma p
\bar{p}$ decays, which was firstly observed~\cite{BESII2GPP} by
BESII, was also confirmed~\cite{BESIIIGammaPP} (see Figure
~\ref{fig:BESIIIgammapp}). This enhancement is fitted with an
$S$-wave Breit-Wigner resonance function, the peak mass is $M =
1861^{+6}_{-13}\mathrm{(stat)}^{+7}_{-26}\mathrm{(syst)} \
\mathrm{MeV/c^2}$ and the width is $\Gamma < 38 \ \mathrm{MeV/c^2}$
at the $90\%$ confidence level. These results are consistent with
published BESII results.

Besides the topics of charmonium physics and light hadron
spectroscopy, many CKM physics related analyses are also carrying on
at BESIII. By exploiting the sample of $\psi(3770)$ decays which is
already the world's largest, some results of the $D$ meson decays
are expected to be published in 2011, including the measurements of
the CKM matrix element $V_{cd}$, the decay constant $f_{D^+}$, and
absolute branching fractions and direct $CP$ violation for some
decay channels.

\section{Data taking plan}
An important aspect of the BESIII running plan is the accumulation
of a sizable $\psi(3770) \rightarrow D\bar{D}$ sample. According to
our assumptions, in two years, i.e. by summer 2011, BESIII will have
a $2.5~ \mathrm{fb^{-1}} \ \psi(3770)$ data sample, three times that
of CLEOc approximately. Furthermore, three years from now, the
current $\psi$' and $J/\psi$ data samples will both be expanded by
nearly an order of magnitude each, and, by the end of the five year
period, we will have a data sample for $D_s$ physics that is few
times than that of the CLEOc sample.

With the few times larger open-charm dataset that will be available
in 2011, BESIII will be able to make better informed decisions on
the physics potential of a very high luminosity $\tau$-charm
factory.

The long-term goals of BESIII include $10~ \mathrm{fb^{-1}}$ samples
at $\psi(3770)$ and at higher $\psi$ states, and a few billion
$J/\psi$ and $\psi$' events. Of course the final sizes of the data
samples will depend on the actual luminosity of BEPCII.

\section{Summary}
Currently, BEPCII is the highest luminosity $e^+e^-$ collider in the
energy region of $\tau$-charm threshold production. A peak
luminosity of $5.6 \times 10^{32} \ \mathrm{cm^{-2} s^{-1}}$ has
been achieved in January 2011. The BESIII detector is also
performing excellently with several physics results already having
been published. In the next few years, a few billion of $J/\psi$,
$\psi$' and approximately $10 \ \mathrm{fb^{-1}}$ of open charm data
will be accumulated. Due to the interesting and rich physics topics
in the charm sector, more experiments are expected to contribute in
this sector. These include the newly operational LHCb experiment,
the approved Super KEK-B/Belle, which will be operational in 2014,
the FAIR/PANDA experiment, planed to be operational in 2015 and the
newly approved Super$B$ super flavor factory at Frascati. In
addition, Novosirbisk has proposed a super-tau-charm factory
project, which may also make substantial contributions to this
field. These proceedings focus on the charm physics programme that
has begun at BESIII, and the future prospects at this experiment.
Note, however that a wide range of important measurements will also
be performed at the Tevatron, at LHCb and the future Super B factory
experiments.

\Acknowledgements

I would like to express my sincere thanks to the whole BESIII
collaboration, the staff of BEPCII and the IHEP computing center. I
would also like to thank the organizers for the wonderful
conference.

\end{document}